\documentclass[amssymb,preprint,preprintnumbers,nofootinbib,superscriptaddress]{revtex4}

\usepackage{graphicx}
\usepackage{latexsym}
\usepackage{amsfonts}
\usepackage{epstopdf}

\begin{document}

\title{Re-visioning the postgraduate preparation of theoretical physicists: An autoethnographic account using the Specialisation dimension of Legitimation Code Theory}

\author{Alan~S.~Cornell}
\email[Email: ]{acornell@uj.ac.za}
\affiliation{Department of Physics, University of Johannesburg, PO Box 524, Auckland Park 2006, South Africa.}

\author{Kershree~Padayachee}
\email[Email: ]{kershree.padayachee@wits.ac.za}
\affiliation{Science Teaching and Learning Unit, Faculty of Science, University of the Witwatersrand, Wits 2050, South Africa.}

\begin{abstract}
There is increasing pressure generally for lecturers to adapt their supervision practices of postgraduate students to better prepare postgraduate students for careers outside of academia. In this paper we examine what such pressure may mean for the supervision and preparation of theoretical physicists specifically, theoretical physics being a sub-discipline of physics usually perceived as a highly specialised niche area of scientific practice. In this exploratory study we apply the concepts of the Specialisation Dimension of Legitimation Code Theory to analyse and reveal the dominant concepts and codes, as well as the code shifts that may occur during postgraduate studies, based on an autoethnographic account of theoretical physicist identity development. The findings demonstrate an underpinning value for both knowledge and knower attributes in the journey to becoming a legitimate theoretical physicist, and the critical role played by postgraduate supervisors in facilitating the process of theoretical physicist identity development. Also highlighted are possible implications for supervisors faced with students intending to take up employment outside of academia.     
\end{abstract}

\maketitle

%
%
\section{Introduction}

\par The universities of today have changed vastly from what they were a century ago. Yet in this age of mass higher education supervision has remained a more personalised form of pedagogy (Bitzer and Albertyn, 2011), suggestive of a time when there was greater socio-cultural homogeneity between university teachers and students, and when the pursuit of a postgraduate qualification in pure sciences generally meant specialisation in a particular field and a life long career in academia. Supervisors enculturated their students in the ways of the discipline (Roseberry, Warren, and Conant, 1992) and students tended to be willing apprentices. Underpinning this enculturation process is the apprenticeship model, premised on the existence of a culture and community of science defined by disciplinary values, models of thinking, patterns of behaviour, language style and an appreciation for a particular ``{\it science lifestyle}" (Lave, 1992; Wang, 2018). Learning, in turn, has been viewed as being the process of developing an identity within this community: 

\par ``{\it Learning is, in this purview, more basically, a process of coming to be, of forging identities in activity in the world. In short, learners are never only that, but are becoming certain sorts of subjects with certain ways of participating in the world$\ldots$ Subjects occupy different locations, and have different interests, reasons and understandings of who they are and what they are up to.}" (Lave, 1992, p. 3)

\par The higher education landscape has, however, shifted dramatically in the past 20 years, both internationally and nationally. Staff and student demographics have changed and the very purpose of higher education has been challenged, with increasing pressure for lecturers to shift postgraduate pedagogical practices to better address the current and future needs of society (Ashwin and Case, 2018). This is particularly true at the doctoral level, where doctoral research and the doctoral qualification is seen as an important means of generating solutions to society's complex problems (CHE, 2014). Academic posts are also increasingly challenging to secure (Roach and Sauermann (2017), and many graduates are forced to look outside academia for employment (Walters, Zarifa and Etmanski, 2020). This branching out of the academic pipeline, with graduates of higher degrees entering workplaces that are vastly different from the academic and research environments for which they have been prepared (Acker and Haque, 2017), indicates the need for doctoral graduates to develop both in-depth and highly sophisticated understandings of the knowledge of their disciplines to be able to apply this knowledge to a broad range of contexts and fields of practice. It may also have implications for the type of identity graduate students develop, based on their intrinsic motivations and their intended career paths. Consequently, doctoral education may need to adapt in response to the shifting societal pressures and requirements to enable the development of PhD graduates with flexible dispositions and a new type of disciplinary ``{\it gaze}" (Bourdieu and Wacquant, 1992) that would allow doctoral graduates to more easily integrate into new contexts and apply their knowledge seamlessly in fields of practice beyond the academe.  This shift may, however, present challenges for the development and training of postgraduates especially in pure, hard sciences, and for PhD supervisors in particular.

\par We seek to interrogate the perceived shift and its implications for doctoral student supervision through an analysis of supervision practices in theoretical physics. Theoretical physics can be considered one of the more abstract, pure sciences, yet despite this, it still needs to become a more professional qualification as the majority of graduates may now pursue careers outside academia, in diverse areas such as banking and industry. For supervisors, this may have ramifications on how the doctoral programme is scaffolded, as the outcomes may need to be revised to reflect a vastly different purpose to what has traditionally been sought. Our purpose is to determine how the preparation of theoretical physicists (and doctoral supervision practices in particular) may need to change whilst still ensuring that the pursuit of obtaining the deep disciplinary knowledge and understanding that is the hallmark of the doctoral qualification, is not compromised.  How does a supervisor prepare students not just for the continued research path of post-doctoral fellowships, but for internships in industry?

\par To pursue this fully we need to first recall that physics, as an academic discipline in science, is historically considered to be one of the hard, pure curricula-oriented type of disciplines (Roberts, 2015). It is underpinned by a hierarchical knowledge structure, strong internal grammar and a strong instructional discourse emphasizing disciplinary knowledge and procedures (Bernstein, 1999; 2000). The undergraduate curriculum especially tends to follow a predetermined structure and sequence, characterised by strong classification and framing (Bernstein, 1999). This curriculum format is, however, necessary to ensure that students become familiar with the language codes and conceptual constellations of physics and are afforded the opportunity to gain the fundamental content knowledge required to progress through the hierarchy of the undergraduate curriculum.   

\par As one advances into the more theoretical branches of physics in advanced years of study, more and more conceptual knowledge must be acquired and integrated. This may leave the experiential approaches to teaching behind, but in learning more concepts and spending more time in lectures, lecturers begin to walk students through an exercise of building piece-by-piece a mental model reflecting the one the lecturer has. This complex process is rarely sufficient to permit students to actually perform genuine problem-solving tasks, as well as integrating them with the range of new procedures they are learning (Middendorf and Pace, 2004). 

\par There is, accordingly, a gradual process of cumulative knowledge building (Maton, 2014) that occurs in the education of physics students, evolving from conceptual knowledge acquisition in the undergraduate setting, to mastery by moving beyond the didactic to understanding and comprehension of the knowledge, and then to the construction of new knowledge at the graduate level. That is, students are gradually inducted into the discipline and the practice of knowledge production, gaining foundational knowledge and `learning to think like a physicist' during undergraduate studies (Van Heuvelen, 1991; Conana, 2016; Conana, Marshall and Case, 2020). Whilst lecturers initially aim to construct disciplinary knowledge in a hierarchical manner, by beginning with hands on experiential work through laboratories and practicals, this quickly becomes a repeated testing of invented threshold concepts (Wisker, 2018) such as Newton's laws etc., to see if these always hold-up under scrutiny, as well as the constant development of students' understanding of complex representations and their uses in physics. Postgraduate study, on the other hand, is focussed on the development of skills for abstracting and producing new knowledge and the gradual development of the theoretical physicist identity.  

\par This repeated testing and questioning that characterises undergraduate studies is a key aspect of the scientist's identity, founded on Dewey's notion of `competent inquiry' (Dewey, 1938, loc. cit. Towne and Shavelson, 2002). It can also be associated with Popper's ``{\it critical rationalism}", amongst other scientific philosophies (Towne and Shavelson, 2002). The key point is that the nature of the scientific laws that are sought to be formulated are abstract ones that must constantly be tested. In the context of theoretical physics, this is captured in the lecture series entitled ``Character of Physical Law", by Feynman (1965), in how he saw ``{\it physical laws}". That is, he suggested a theoretical physicist attempts to write down, in the language of mathematics, a simple and beautiful explanation for nature. However, he does not impose this on nature. It is at best a model, one that will only have a certain regime of applicability. Physicists test the range of this model, whilst refining and improving, in a search for new laws and broader or deeper understandings and applications, making this the cornerstone of physics education. 

\par The inquiry mind set, while triggered in undergraduate study, must be fostered more intentionally during the crucial transitional time of doctoral studies in particular, a time which has been considered the time of the academic's identity construction or a rite-of-passage to becoming an academic or expert in the field. As highlighted by Wisker (2018), the research learning when doctoral candidates make breakthroughs in their thinking, understanding, researching and writing largely concern conceptual threshold crossing (Meyer and Land, 2003), which show both ontological change (changing the way they see themselves in the world and in their identity as a researcher) and epistemological change (a confidence in engaging with the research learning, and an active awareness of the ways of constructing knowledge and making a contribution).

\par While the knowledge outcomes may be clearly identified for students, the cultural norms and values (set by the research scientists (Brickhouse and Schultz, 1999), that enable genuine participation in the discipline) are usually less well defined. This may also be one of the factors contributing to the relatively high attrition rate of students in undergraduate physics programmes and the reason for the ``{\it pipeline metaphor}" which ``{\it models physics retention as a stream of students flowing through a physics pipeline}" until they ``{\it leak}" out (leaving physics) or arrive at a fixed endpoint where they are ``{\it full-fledged physicists}" (Quan 2017, p 16).

\par In this paper, we attempt to make the tacit more explicit by drawing on the personal experiences and insights of one of the authors, an academic in this discipline, to highlight the iterative cycles of epistemological and ontological shifts that occur in the process of theoretical physicist identity development as students move from novice to developing expert and finally to expert. Using concepts from the Specialisation Dimension of Legitimation Code Theory (LCT) (Maton, 2014), we demonstrate what is valued in the discipline at different stages of undergraduate and postgraduate education, specifically from the supervisor perspective, and explain how these values influence the development of the identity and ``{\it cultivated gaze}" (Maton, 2014) of an expert theoretical physicist. We expose the organising principles that shape the disciplinary education of theoretical physicists, and we reflect on what, if anything, might be required to prepare PhD graduates for different fields of practice. 

%
%
\section{Application of the Specialisation Dimension of LCT and Science Identity Theory}

\par LCT is a multidimensional conceptual toolkit for illuminating struggles of legitimacy in social fields of practice, allowing for ``{\it both the exploration of knowledge-building and the cumulative building of knowledge}", and it  ``{\it enables knowledge practices to be seen, their organizing principles to be conceptualized, and their effects to be explored}" (Maton, 2014, p. 4). It is comprised of four dimensions: Autonomy, Specialisation, Semantics and Temporality (Maton, 2014), ``{\it each dimension comprises a series of concepts centred on capturing a set of organizing principles underlying dispositions, practices and contexts}" (Maton, 2016, p. 11). 

\par The Specialisation dimension of LCT enables the exploration of practices in terms of knowledge-knower structures (Maton, 2014). The organising principles of specialisation may be described in terms of {\it epistemic relations} (ER) and {\it social relations} (SR). Epistemic relations (ER) refer to the strength of the relationship between a practice and its object; Social relations (SR) refer to the strength of the relationship between practices and their subject (Maton, 2014). The relative strengths and weaknesses of these relations may be indicated by ``+" and ``--", respectively. These relations may be combined to illustrate the continuum of strengths, leading to the creation of the Specialisation plane and four Specialisation codes, viz., Knowledge Codes, Knower Codes, Elite Codes and Relativist Codes (Maton, 2014) (Fig. 1).

\begin{figure}
\includegraphics[width=0.6\textwidth]{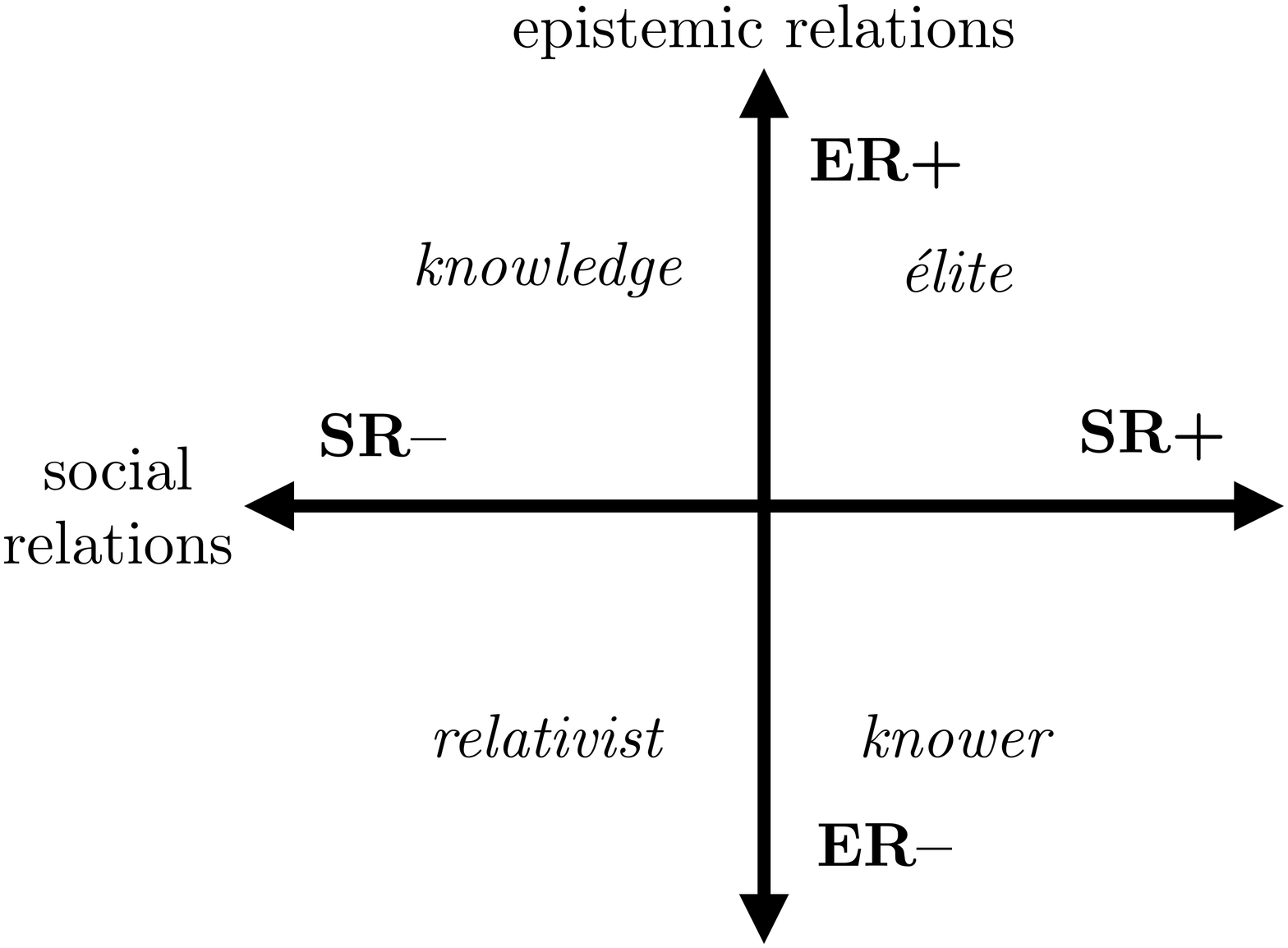}
\caption{The Specialisation plane of LCT, showing the four specialisation codes.}\label{Fig:1}
\end{figure}

\noindent Maton (2014, p. 31) describes Specialisation codes within the Specialisation plane as follows:
\begin{itemize}
\item Knowledge codes (ER+, SR--), ``$\ldots$ {\it where possession of specialised knowledge, principles, or procedures concerning specific objects of study is emphasised as the basis of achievement, and the attributes of actors are downplayed}"; 
\item Knower codes (ER--, SR+), ``$\ldots$ {\it where specialised knowledge and objects are downplayed and the attributes of actors are emphasised as measures of achievement}" 
\item \'Elite codes (ER+, SR+), ``$\ldots$ {\it where legitimacy is based on both possessing specialist knowledge and being the right kind of knower}"; and 
\item Relativist codes (ER--, SR--), ``$\ldots$ {\it where legitimacy is determined by neither specialist knowledge nor knower attributes -- `anything goes}'". 
\end{itemize}

\par Given the focus of the study, the concepts and codes of specialisation thus provide the language and tools to be able to identify the logics that underpin the education and identity development of theoretical physicists. However, LCT also provide other tools, such as the 4-K model of epistemic relations, that may be employed to reveal the specific knowledge and practices which are valued in a discipline, as well as what is viewed as acceptable objects of study. The 4-K model distinguishes epistemic relations into two further categories: ontic relations (OR), describing the strengths of relations between knowledge practices and their objects, and discursive relations (DR), referring to the strengths of relation between knowledge practices and other knowledge practices (Maton, 2014).  

\par While LCT provides a useful set of tools for revealing the underpinning logics that shape supervision practices, the findings must still be viewed in relation to identity development to determine the implications for supervision practices. For this purpose, we refer to the ``{\it science identity}" model proposed by Carlone and Johnson (2007) as the overarching theoretical framework for this study. In this model (informed by the work of Gee (2000)), science identity develops as a consequence of a complex interplay between competency, performance and recognition (Fig. 2). Competence refers to the knowledge and understanding of scientific content, performance is the ability to enact or demonstrate the practices of science to others, while recognition refers to the self-recognition or recognition by others as a scientist. According to the science identity model, science identity is shaped as a consequence of varying levels of development in these three broad areas, bearing in mind that both personal agency as well as various contextual factors may also influence the process. Carlone and Johnson highlight, for instance, that gender, racial and ethnic identities may affect science identity, but also that an identity can only exist to that extent one is able to demonstrate that identity in a way that is recognisable and can be validated by others. Science identity, according to this model, is thus, ``{\it both situationally emergent and potentially enduring over time and context}" (Carlone and Johnson, 2007, p. 1192). It should be noted that for the purposes of this study, our focus is on how these three areas (i.e., competence performance and recognition) are influenced and developed by supervision practices and the supervisor's own science identity. We acknowledge the influence of agency and other contextual factors such as race or socio-economic background, however, it is beyond the scope of the present exploratory study. It will however, be integrated into a wider study on supervision practices and identity in Science Technology Engineering and Mathematics (STEM) disciplines, which this study serves as a precursor for.  

\begin{figure}
\includegraphics[width=0.6\textwidth]{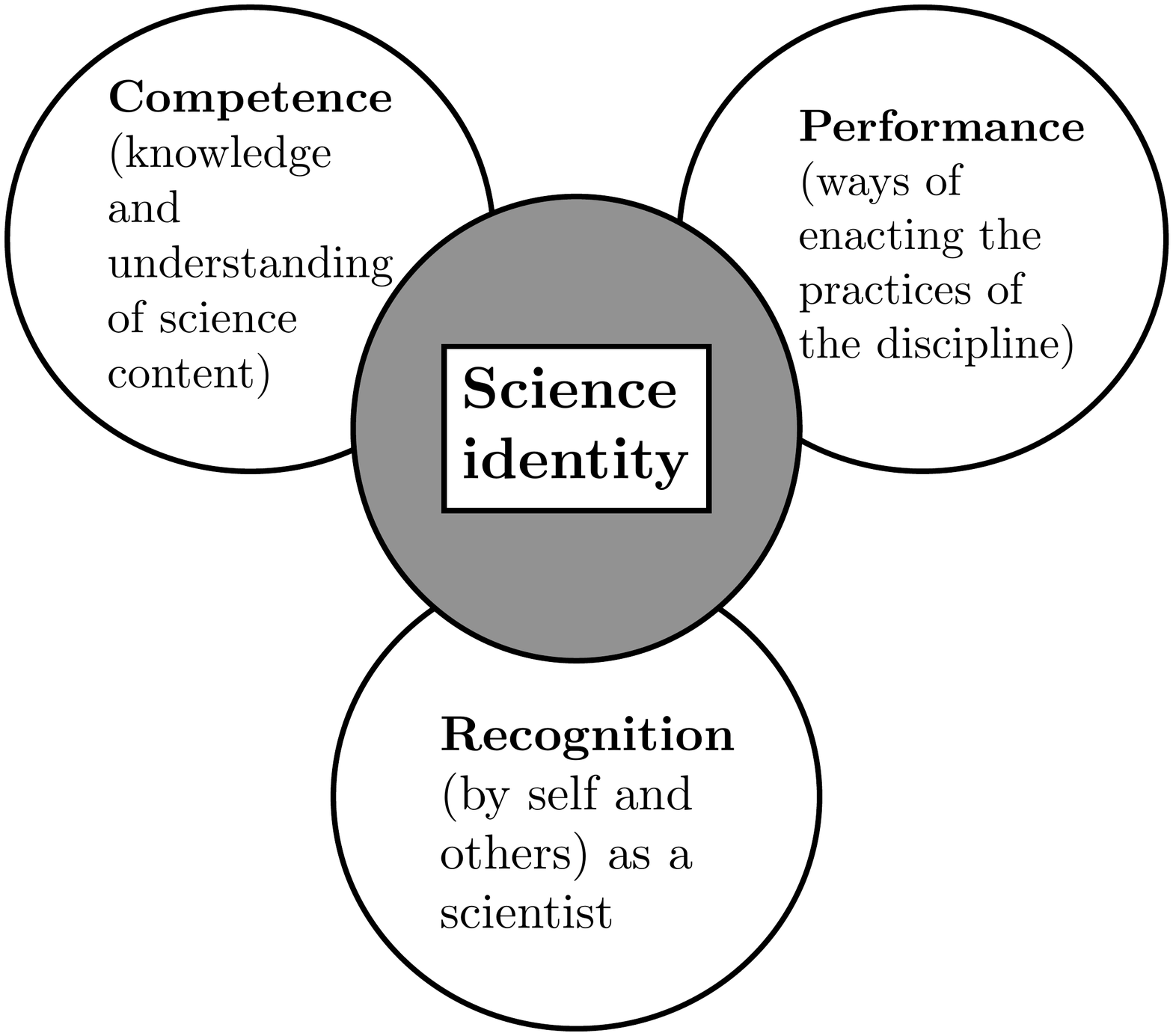}
\caption{Model of Science Identity (adapted from Carlone and Johnson (2007)).}\label{Fig:2}
\end{figure}

%
%
\section{Methodology}

\par This study followed a qualitative research paradigm, using a facilitated autoethnographic research method (Hamilton, Smith and Worthington, 2008) for data generation. Data was drawn from a semi-structured interview facilitated by author 2, a science education specialist, in which author 1, a theoretical physicist specialising in high energy particle physics and gravitational physics, reflected on his own developmental journey and his current postgraduate supervision practices. We were aware of the limitations of the autoethnographic approach, however, our intention was to provide an in depth account of an individual experience of moving through a particular culture, in this case the culture of theoretical physics. As both Stephen, O'Connor and Garrison (2005) and Hamilton et al. (2008) argue, methodological approaches like autoethnographies can illustrate how small, sometimes idiosyncratic, experiences can have a cascading effect in students' broader trajectories. Therefore, considering the focus of this research, facilitated autoethnography was determined to be the most appropriate method, despite the limitations. Credibility was, however, enhanced to an extent, by conducting a facilitated autoethnography, while trustworthiness was addressed through the recording, transcription and subsequent verification of the interview.

\par The interview transcript was initially examined broadly to identify emergent themes and the dominant specialisation concepts within the data. This initial coarse analysis was followed by a more fine grained analysis using a translation device designed specifically to identify the manifestations of these concepts as well as the different strengths of these concepts. The strengths of the concepts ranged from strong and very strong epistemic relations (ER+ and ER++ respectively), to weak and very weak epistemic relations (ER-- and ER-- --). The continuum of strengths of social relations was similarly classified as strong and very strong social relations (SR+ and SR++), and weak and very weak social relations (SR-- and SR-- --).  The data emerging from this analysis were plotted on the specialisation plane (Fig. 3), with spread of the data points highlighting the code shifts that a theoretical physicist may experience in the progression from undergraduate studies to PhD. The data were also examined for evidence of shifts in epistemic relations to explicate the development of different gazes and insights. 

%
%
\section{Results and Discussion}

\par The clustering of data points in the knowledge and elite quadrants of the specialisation plane (Fig. 3) confirmed the dominance of the knowledge code in theoretical physics, but also indicated a distinct code shift in the transition from undergraduate student to expert in this discipline. For instance, the clustering of points related to matriculants (school leavers) applying to study physics, in the knowledge quadrant demonstrates the dominance of strong epistemic relations even at this initial stage, with applicants' mathematics and science marks being viewed as an indication of an applicant's foundational knowledge and understanding (i.e., foundational competence). The strong value for competence as a critical aspect of a nascent theoretical physicist identity was also evident in the clustering of data points related to undergraduate studies in the knowledge quadrant of the specialisation plane. The pattern confirmed that at this stage of identity development, knowledge acquisition, familiarisation of students with disciplinary concepts, and the gradual development of epistemological constellations is foregrounded. 

\begin{figure}
\includegraphics[width=0.6\textwidth]{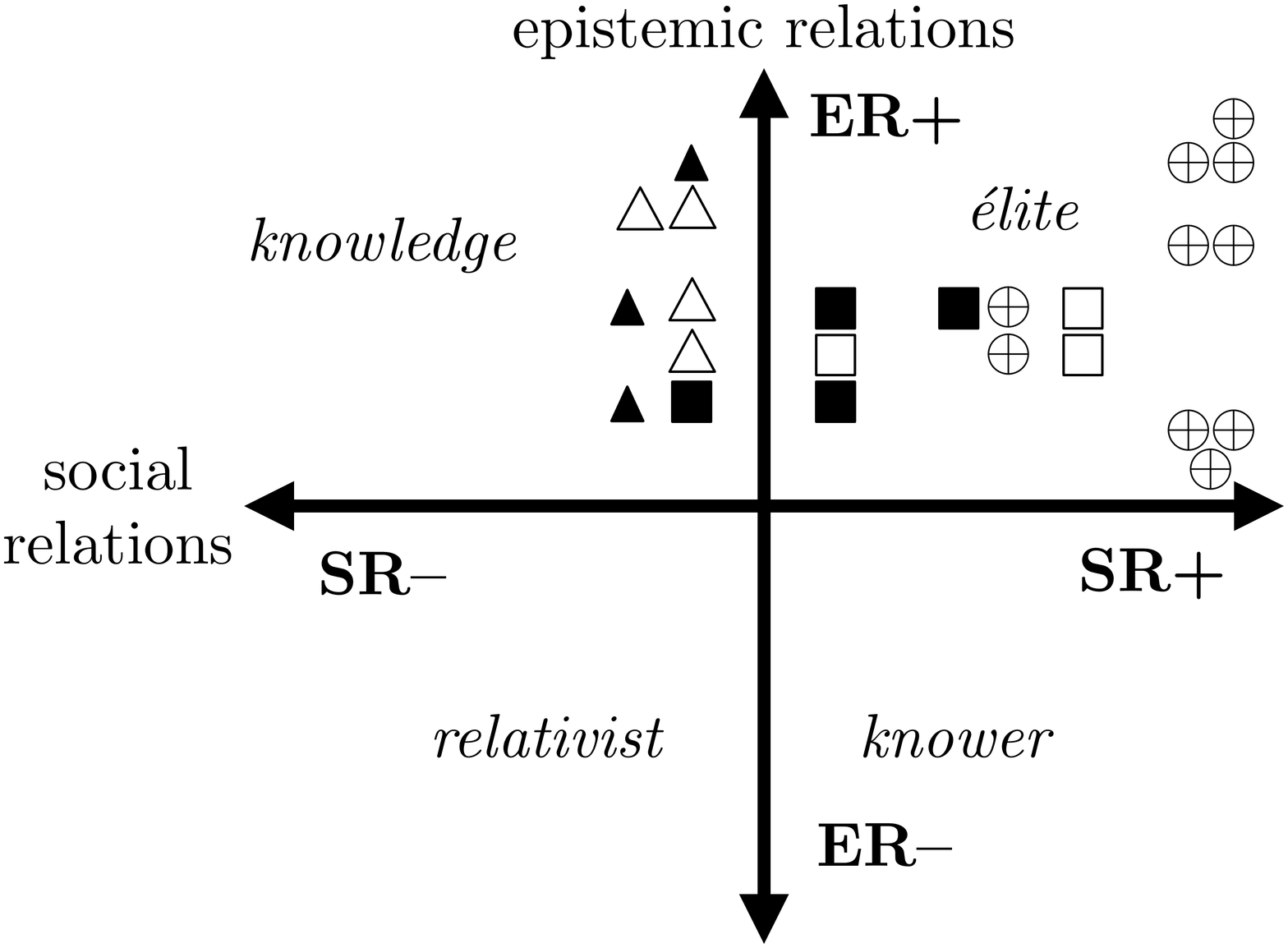}
\caption{The data points $\blacktriangle$ refer to a matriculant entering varsity, $\triangle$ refers to undergraduate education/pedagogies, $\blacksquare$ to honour's, $\square$ Masters and PhD,  to post-doctoral, and $\oplus$ to an expert.}\label{Fig:3}
\end{figure}

\par The endurance of the knowledge of physics and the notion that there is a particular ``{\it right way}" to engage with this knowledge (as highlighted in the quote below), is particularly noteworthy: 

\par ``{\it Pedagogies are very often traditional. I think there are a few places trying new things, you know, like flipped classrooms, but it's still very much the same content as probably the content of the last 100 years. The textbook from 1952 is still the same textbook content-wise, as was used in 2012. Because of the nature of physics being so hierarchical, there's not really much can that can be changed. So, it's largely presented the same way too.}" 

\par The idea of ``right answers" and ``right procedures", as well as the strong classification and framing highlighted in the above quote, are indicative of both strong ontic relations (OR+) suggesting high ontic fidelity, as well as strong discursive relations and procedural purism (Maton, 2014). In this respect, undergraduate study is characterized by strict rules applied in tightly bound contexts, with the lecturer (as the expert) setting the rules for the engagement with the knowledge. However, it also confirms the importance of competence in the identity of a theoretical physicist. Performance development is also addressed to some extent in the course of undergraduate studies, with students entering university studies in physics faced with time intensive lectures and laboratories designed to build their conceptual knowledge base and to develop an appreciation for the methodology of basic sciences.

\par As students progress through the undergraduate qualification, competence and performance is developed through the deepening of conceptual knowledge and the integration of more complex procedures and practices, with emphasis shifting to problem solving and broader application of concepts: 

\par ``{\it Problem solving becomes more important towards 3rd year. There's a focus on problems outside of the box -- problems you can't fit into lectures.}"

\par Students may thus, experience epistemological shifts through the undergraduate curriculum, however, ontological shifts associated with identity development may not be guaranteed. In this regard, there appeared to be tacit acknowledgement that ontological shifts require a certain pre-existing disposition, referred to in the interview as ``{\it an intuition possessed by some students}" that is indicative of their potential to become fully fledged theoretical physicists. This ``{\it intuition}" is suggestive of knower attributes and the possibility of a born gaze (Maton, 2014) being an important component in overall theoretical physicist identity development. Given the strong emphasis of the dominance of the knowledge code in the undergraduate curriculum, acknowledgement of particular knower attributes as an important feature at this early stage of identity development was somewhat unexpected, and perhaps not often interrogated in the traditional teaching pedagogies, given the focus on traditional, transmission-based pedagogies mentioned earlier. 

\par On graduation, the data suggest the expectation that students will have developed a ``{\it trained}" gaze (Maton, 2014), having achieved sufficient conceptual grounding (competence) and familiarity with the practices of physics (performance) to enable a shift to a more sophisticated way of engaging with the knowledge of the discipline in postgraduate studies. However, this trained gaze (Maton, 2014) is still considered insufficient for legitimate peripheral participation (Lave and Wenger, 1991), as indicated below:

\par ``{\it There is so much you have to learn before you get to that point (of being recognized as an expert). You have to have a certain command of a lot of the language and procedures}$\ldots$"

\par It is the focussed development of a certain command of the discourse of theoretical physics that is emphasised at postgraduate level, coupled with induction into the socio-cultural practices of the discipline (where the culture and disciplinary ``{\it norms}" become visible), that facilitate the emergence of the cultivated gaze (Maton, 2014) required for legitimate participation. 

\par The data shows that the real shift signalling a move towards identifying as a legitimate theoretical physicist only really appears at PhD and postdoctoral levels. If a student has developed a sufficiently deep foundational knowledge base, they will be able to not only apply that knowledge but start to evolve the knowledge by conceptualising applications of knowledge in different, possibly unrelated contexts. An equally important shift that occurs at this stage is the strengthening of social relations, where communication of ideas and networking becomes important and where the role of the supervisor is especially critical. In the author's account of his own developmental journey, this is highlighted by the following quotes from the interview: 

\par ``{\it The supervisor is seen as the expert in the discipline. The PhD is the equivalent of the apprenticeship.}" 

\par ``{\it During my PhD, I produced 5 research papers. The first few, the supervisor conceives the idea, the last few, the supervisor shifts the work of conceptualization to the student as well.}"

\par ``{\it Emphasis on postdoc studies -- produce papers. This shows the ability to publish but also shows the knowledge and skills. But references are also important -- who you know -- your network that you build during multiple postdocs (and big collaborators) is really important.}"

\par The last quote above signals the importance of recognition in the identity of a theoretical physicist, and the pivotal roles of publications and networking. In this respect, the supervisor plays the primary role in facilitating these first connections in forming a collaborative network, and will often assist in finding their first academic and research positions. The tools and connections the supervisor provides therefore assist the student in their post-PhD experiences, as they build new connections of their own, through attendances at conferences and workshops. Provision of such opportunities by PhD supervisors and the individual's own engagement in multiple postdoctoral fellowships thereafter, are thus, key factors in identity development. These experiences facilitate legitimacy through publications (the impact and renown that these publications have, viewed as strong signals of ability to perform independently as a theoretical physicist) while the post-doctoral fellowships enable the development of a network of colleagues. This demonstrates the strong social relations that dominate at this stage of development, as one moves closer to being identified as a theoretical physicist. It does, however, highlight that identity is very strongly linked to interactions with others in communities of practice (Lave and Wenger, 2001), and access and integration into the academic tribe is facilitated largely through competence and performance:  

\par ``{\it If I meet someone at a conference, you start chatting with this person, you can tell almost immediately if they have a certain confidence within themselves about the knowledge they're talking about. Instead of being a bit vague, they have a particular outlook. They have developed their own ideas and their own gravitas towards it. The way he or she conducts themselves or asks questions signals that they know what they're talking about.}"

\par The `outlook' mentioned in the above quote may be equated to the cultivated gaze of an expert in the discipline, which is as much dependent on recognition as is it on competency and performance, with the undergraduate and early postgraduate years of study focused on the latter, and recognition developed more extensively in latter years of PhD and postdoctoral studies.   

\par The integrative role of the supervisor at PhD level is thus very significant in developing the cultivated gaze of a theoretical physicist, the model of supervision (Lee, 2007) tending towards critical thinking with elements of enculturation. As has been explained, this is due to the hierarchical structure of the discipline which requires an immersion in the background theory before ``{\it problematising}" and then ``{\it finding connections}" (Lee, 2007). However, such a hierarchy emphasises a one-sided power relationship, where the supervisor decides what is legitimate knowledge (Bernstein, 1999). Also highlighted is that identity of emergent theoretical physicists is very strongly linked to the identity and values of the supervisor, and a significant part of this identity is framed in terms of recognition and social relations. This does then suggest that there may indeed be some tension for supervisors in enabling recognition in circles outside of academia. However, as mentioned by Carlone and Johnson, identity is emergent, and if the development of the other elements of identity, viz., competence and performance are well supported, students may be able to shift from a purist to a doctrinal insight (Maton, 2014) when required to apply themselves in a different context. 

%
%
\section{Concluding remarks}

\par Science identity formation is a complex phenomenon arising from the interactions between an individual and a particular body of scientific knowledge as well as the individual and other scientists. For theoretical physics, identity is initially shaped by strong interactions with the knowledge and procedures of the discipline, with emphasis placed on the development of the competence and performance aspects of science identity. However, knower attributes and social relations play an increasingly important role in the transition from learning about physics to becoming and being recognised as a theoretical physicist. For students and graduates on a career path to academia, this transition may be more easily facilitated by supervisors who are themselves recognised as legitimate theoretical physicists. For students intending to enter workplaces outside of academia, however, the process of recognition building may be more challenging both for the students and their supervisors. As pointed out by Carlone and Johnson (2007), identity is shaped by the contexts in which one operates. It is, therefore, possible that the strong knowledge competence and performance attributes developed during undergraduate and postgraduate studies may be leveraged by those students to gain recognition of a different type in different contexts. This possibility, as well as other ways that STEM supervisors might be envisaging shifting supervision practices to better prepare graduates for diverse working environments, will be explored further in a related study.

%
%

\section*{Acknowledgements}

\noindent ASC is supported in part by the National Research Foundation of South Africa (NRF).  

%
%
\section*{References}

\noindent Acker, S., and Haque, E. (2017). Left out in the academic field: Doctoral graduates deal with a decade of disappearing jobs. {\it Canadian Journal of Higher Education/Revue canadienne d'enseignement sup\'erieur}, 47(3), 101-119.\\
Ashwin, P., and Case, J. M. (2018). {\it Higher Education Pathways: South African Undergraduate Education and the Public Good} (p. 308). African Minds.\\
Bernstein, B. (1999) Vertical and horizontal discourse: An essay. {\it British Journal of Sociology and Education} 20: 157-73.\\
Bernstein, B. (2000). {\it Pedagogy, symbolic control, and identity: Theory, research, critique} (Vol. 5). Rowman and Littlefield.\\
Carlone, H. B., and Johnson, A. (2007). Understanding the science experiences of successful women of color: Science identity as an analytic lens. {\it Journal of Research in Science Teaching: The Official Journal of the National Association for Research in Science Teaching, 44}(8), 1187-1218.\\
Council on Higher Education (2014). Vital stats: public higher education 2014. Pretoria: Council on Higher Education.\\
Conana, C. H. (2016). Using semantic profiling to characterize pedagogical practices and student learning: A case study in two introductory physics courses.\\
Conana, H., Marshall, D., and Case, J. (2020). A SEMANTICS ANALYSIS OF FIRST-YEAR PHYSICS TEACHING. {\it Building Knowledge in Higher Education: Enhancing Teaching and Learning with Legitimation Code Theory}.\\
Feynman, R P (1965) The Character of Physical Law, BBC.\\
Gee, J. P. (2000). Chapter 3: Identity as an analytic lens for research in education. {\it Review of research in education, 25}(1), 99-125.\\
Lave, J., and Wenger, E. (1991). {\it Situated learning: Legitimate peripheral participation.} Cambridge university press.\\
Lave, J. (1992). Word problems: A microcosm of theories of learning. {\it Context and cognition: Ways of learning and knowing}, 74-92.\\
Lee, A. M. (2007). Developing effective supervisors: Concepts of research supervision. {\it South African Journal of Higher Education, 21}(4), 680-693.\\
Margolis, E. (Ed.). (2001). {\it The hidden curriculum in higher education}. Psychology Press.\\
Maton, K. (2014) {\it Knowledge and Knowers: Towards a realist sociology of education}, London: Routledge.\\
Maton, K. (2016). Legitimation Code Theory: Building knowledge about knowledge-building.\\
Meyer, J., and Land, R. (2003). {\it Threshold concepts and troublesome knowledge: Linkages to ways of thinking and practising within the disciplines} (pp. 412-424). Edinburgh: University of Edinburgh.\\
Middendorf, J., and Pace, D. (2004). Decoding the disciplines: A model for helping students learn disciplinary ways of thinking. {\it New directions for teaching and learning, 2004}(98), 1-12.\\
Quan, G. (2017). {\it Becoming a Physicist: How Identities and Practices Shape Physics Trajectories} (Doctoral dissertation).\\
Roach, M., and Sauermann, H. (2017). The declining interest in an academic career. {\it PLoS One, 12}(9), e0184130.\\
Rosebery, A. S., Warren, B., and Conant, F. R. (1992). Appropriating scientific discourse: Findings from language minority classrooms. {\it The Journal of the Learning Sciences, 2}(1), 61-94.\\
Roberts, P. (2015) Higher education curriculum orientations and the implications for institutional curriculum change. {\it Teaching in Higher Education}, vol. 20, no. 5, 542-555. \\
Stevens, R., O'Connor, K., and Garrison, L. (2005). Engineering student identities in the navigation of the undergraduate curriculum. In {\it Proceedings of American Society for Engineering Education Annual Conference} (pp. 1-8).\\
Towne, L., and Shavelson, R. J. (2002). {\it Scientific research in education}. National Academy Press Publications Sales Office.\\
Van Heuvelen, A. (1991). Learning to think like a physicist: A review of research?based instructional strategies. {\it American Journal of physics}, 59(10), 891-897.\\
Walters, D., Zarifa, D. and Etmanski, B. Employment in Academia: To What Extent Are Recent Doctoral Graduates of Various Fields of Study Obtaining Permanent Versus Temporary Academic Jobs in Canada? {\it High Educ Policy} (2020). https://doi.org/10.1057/s41307-020-00179-w\\
Wang, C. (2018). Scientific Culture and the Construction of a World Leader in Science and Technology. {\it Cultures of Science}, 1(1), 1-13.\\
Wisker (2018) ``Different Journeys: Supervisor Perspectives on Disciplinary Conceptual Threshold Crossing in Doctoral Learning", {\it CriSTaL} vol 6, issue 2.

\end{document}